\documentclass[aps,prl,twocolumn,floatfix,superscriptaddress,showpacs,amsfonts,amssymb,amsmath,preprintnumbers]{revtex4}
\usepackage{bm}
\usepackage{psfig}
\usepackage{dcolumn}

\newcommand{\Eqref}[1]{Equation~(\ref{#1})}
\renewcommand{\eqref}[1]{Eq.~(\ref{#1})}

\newcommand{\Eqsdash}[2]{Equations~(\ref{#1}--\ref{#2})}

\newcommand{\exsdash}[2]{(\ref{#1}--\ref{#2})}

\newcommand{\figref}[1]{Fig.~\ref{#1}}

\newcommand{\bea}{\begin{eqnarray}}
\newcommand{\eea}{\end{eqnarray}}
\newcommand{\bal}{\begin{aligned}}
\newcommand{\eal}{\end{aligned}}
\newcommand{\bga}{\begin{gathered}}
\newcommand{\ega}{\end{gathered}}

\newcommand{\bl}{\bigl}
\newcommand{\br}{\bigr}
\newcommand{\la}{\langle}
\newcommand{\ra}{\rangle}

\renewcommand{\phi}{\varphi}

\newcommand{\dd}{\partial}
\newcommand{\diff}{d}

\newcommand{\vu}{{\bf u}}
\newcommand{\vf}{{\bf f}}
\newcommand{\vB}{{\bf B}}
\newcommand{\tB}{{\tilde B}}
\newcommand{\vb}{{\skew{-4}\hat{\bf b}}}
\newcommand{\bb}{{\hat b}}
\newcommand{\vx}{{\bf x}}

\newcommand{\vk}{{\bf k}}
\newcommand{\tvk}{{\tilde \vk}}
\newcommand{\tk}{{\tilde k}}
\newcommand{\vn}{{\hat{\bf k}}}
\newcommand{\nn}{{\hat k}}

\newcommand{\kperp}{k_\perp}
\newcommand{\kpar}{k_\parallel}
\newcommand{\kapperp}{\kappa_\perp}
\newcommand{\kappar}{\kappa_\parallel}
\newcommand{\Iperp}{I_\perp}
\newcommand{\Ipar}{I_\parallel}
\newcommand{\Cone}{C_{\perp\perp}}
\newcommand{\Cperp}{C_{\parallel\perp}}
\newcommand{\Cpar}{C_{\parallel\parallel}}

\newcommand{\Ii}{I^\text{(i)}}
\newcommand{\Ia}{I^\text{(a)}}
\newcommand{\Ea}{\tilde E}
\newcommand{\kapI}{\kappa^\text{(i)}}
\newcommand{\kapA}{\kappa^\text{(a)}}

\newcommand{\kf}{k_0}
\newcommand{\ks}{k_\text{s}}
\newcommand{\kd}{k_\nu}
\newcommand{\kres}{k_\eta}

\newcommand{\lf}{\ell_0}
\newcommand{\ls}{\ell_\text{s}}
\newcommand{\ld}{\ell_\nu}
\newcommand{\lres}{\ell_\eta}

\newcommand{\tcorr}{\tau_\text{c}}
\newcommand{\rK}{r_\text{2D}}
\newcommand{\CK}{C_\text{K}}

\newcommand{\Wo}{W_0}
\newcommand{\Wd}{W_\nu} 

\newcommand{\urms}{u_\text{rms}}

\newcommand{\usq}{\la u^2\ra}
\newcommand{\Bsq}{\la B^2\ra}

\renewcommand{\Pr}{\text{Pr}_\text{m}} 
 
\renewcommand{\Re}{\text{Re}} 
\newcommand{\Rel}{\text{Re}_\lambda}

\newcommand{\gKA}{{\bar \gamma}}
\newcommand{\gone}{\gamma_{\perp}}

\newcommand{\sperp}{\sigma_\perp}
\newcommand{\spar}{\sigma_\parallel}

\begin{document}

\preprint{{\em Phys.\ Rev.\ Lett.}\ {\bf 92}, 084504 (2004); 
e-print {\tt astro-ph/0308252}}

\title{Saturated State of the Nonlinear Small-Scale Dynamo}

\author{A.\ A.\ Schekochihin}
\email{as629@damtp.cam.ac.uk}
\altaffiliation[present-time address: ]{DAMTP/CMS, 
University of Cambridge, Wilberforce Road, Cambridge CB3 0WA, UK.}
\affiliation{Plasma Physics Group, Imperial College, 
Blackett Laboratory, Prince Consort Road, London~SW7~2BW, UK}
\author{S.\ C.\ Cowley}
\affiliation{Plasma Physics Group, Imperial College, 
Blackett Laboratory, Prince Consort Road, London~SW7~2BW, UK}
\affiliation{Department of Physics and Astronomy, 
UCLA, Los Angeles, CA 90095-1547}
\author{S.\ F.\ Taylor}
\affiliation{Plasma Physics Group, Imperial College, 
Blackett Laboratory, Prince Consort Road, London~SW7~2BW, UK}
\author{G.\ W.\ Hammett}
\affiliation{Plasma Physics Laboratory, Princeton University, 
Princeton, NJ 08543}
\author{J.\ L.\ Maron}
\affiliation{Department of Physics and Astronomy, University of Rochester, 
Rochester, NY 14627} 
\affiliation{Department of Physics and Astronomy, University of Iowa, 
Iowa City, IA 52242}
\author{J.\ C.\ McWilliams}
\affiliation{Department of Atmospheric Sciences, 
UCLA, Los Angeles, CA 90095-1565}
\date{\today}

\begin{abstract}
We consider the problem of incompressible, forced, nonhelical, homogeneous 
and isotropic MHD turbulence with no mean magnetic field 
and large magnetic Prandtl number. 
This type of MHD turbulence 
is the end state of the turbulent dynamo, which generates folded 
fields with small-scale direction reversals. 
We propose a model in which 
saturation is achieved as a result of the velocity statistics 
becoming anisotropic with respect to the local direction 
of the magnetic folds. The model combines the effects of 
weakened stretching and quasi-two-dimensional mixing 
and produces magnetic-energy spectra 
in remarkable agreement with numerical results 
at least in the case of a one-scale flow. 
We conjecture that the statistics seen in numerical simulations 
could be explained as a superposition of 
these folded fields and Alfv\'en-like waves that propagate along the folds. 
\end{abstract}


\maketitle

In this Letter, we consider what is perhaps 
the oldest formulation of the MHD turbulence problem dating back 
to Batchelor's work in 1950~\cite{Batchelor_dynamo}: 
incompressible, randomly forced, nonhelical, homogeneous, isotropic 
MHD turbulence described by 
\bea
\label{NSEq}
&&\dd_t\vu + \vu\cdot\nabla\vu =
\nu\Delta\vu - \nabla p + \vB\cdot\nabla\vB + \vf,\\
&&\label{ind_eq}
\dd_t\vB + \vu\cdot\nabla\vB = 
\vB\cdot\nabla\vu + \eta\Delta\vB.
\eea
The pressure~$p$ (determined from $\nabla\cdot\vu=0$)
and the magnetic field~$\vB$ are 
rescaled by~$\rho$ and~$(4\pi\rho)^{1/2}$, respectively 
($\rho$ is density). 
Turbulence is excited by the random external forcing~$\vf$. 
No mean field is imposed. 
We are primarily interested 
in the case of the large magnetic Prandtl number~$\Pr=\nu/\eta$ 
which is appropriate for the warm interstellar medium 
and cluster plasmas~\cite{Widrow_review}. 
Numerical 
evidence suggests that the popular choice~$\Pr=1$ is 
in many respects similar to the large-$\Pr$ regime~\cite{SCTMM_stokes}. 
$\Pr\gg1$ implies 
that the resistive scale~$\lres\sim\Pr^{-1/2}\ld$ is much smaller than 
the viscous scale~$\ld$. 
Thus, the problem has two scale ranges: the hydrodynamic (Kolmogorov) 
inertial range $\lf\gg\ell\gg\ld\sim\Re^{-3/4}\lf$ 
($\lf$ is the forcing scale) 
and the subviscous range $\ld\gg\ell\gg\lres$. 

For a moment, let us consider the traditional view of fully developed 
incompressible MHD turbulence in the presence of a strong, externally 
imposed mean field. This view is based on the idea of 
Iroshnikov~\cite{Iroshnikov} and 
Kraichnan~\cite{Kraichnan_MHD} that it is a turbulence 
of strongly interacting Alfv\'en-wave packets. 
This phenomenology, modified by Goldreich and Sridhar~\cite{GS_strong} 
to account for the anisotropy induced by 
the mean field, predicts steady-state spectra for magnetic 
and kinetic energies that are identical in the inertial range and 
have Kolmogorov $k^{-5/3}$ scaling. 
An essential feature of this 
description is that it implies scale-by-scale 
equipartition between magnetic and velocity fields: indeed, 
$\delta\vu_\vk=\delta\vB_\vk$ in an Alfv\'en wave. 
Simulations appear to confirm Alfv\'enic equipartition 
provided there is an imposed strong 
mean field $B_0\gg\urms$~\cite{Maron_Goldreich}. 

\begin{figure}[t]
\centerline{\psfig{file=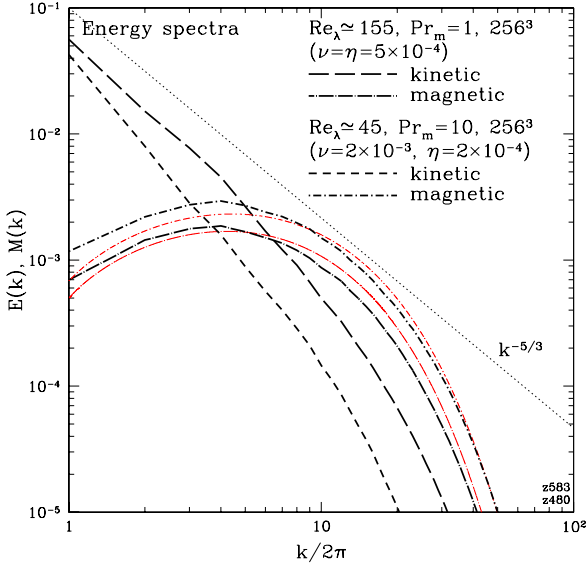,width=8cm}}
\caption{\label{fig_AB} 
Energy spectra in simulations with $\Pr=1$, $\Rel\simeq155$ 
and with $\Pr=10$, $\Rel\simeq45$ (bold lines). 
The thin lines are our model-predicted spectra 
of the folded field component (normalized to have 
the same energy as the numerical spectra).} 
\end{figure}

In the case of zero mean field, it has been widely assumed 
that essentially the same description applies, except it is 
the large-scale magnetic fluctuations that 
play the role of effective mean field along which smaller-scale Alfv\'en waves 
can propagate. However, 
numerical simulations of isotropic MHD turbulence do not show 
scale-by-scale equipartition between kinetic and magnetic energies.
There is a definite and very significant excess of 
magnetic energy at small scales. This is true both for $\Pr>1$ and 
$\Pr=1$ (\figref{fig_AB}). This result persists at 
the highest currently available 
resolution ($1024^3$, see~\cite{Haugen_Brandenburg_Dobler}).

Let us consider the genesis of the 
magnetic field in isotropic MHD turbulence. As there is no mean field, 
all magnetic fields are fluctuations 
generated by the small-scale dynamo. This type of dynamo 
is a fundamental mechanism that amplifies magnetic energy in 
chaotic 3D flows with sufficiently large magnetic Reynolds numbers 
and~$\Pr\gtrsim1$. 
The amplification is due to random stretching of the 
magnetic-field lines by the velocity field.
During the kinematic (weak-field) stage of the dynamo,  
the magnetic energy grows exponentially in time, its 
spectrum is peaked at the resistive scale, $\kres\sim\Pr^{1/2}\kd$, 
and grows self-similarly~\cite{Kazantsev,KA,SCTMM_stokes}. 
The growth rate is of the order of the turnover rate 
of the fastest eddies, which, in Kolmogorov 
turbulence, are the viscous-scale ones. 

Although the bulk of the magnetic energy 
is at the resistive scale, the dynamo-generated fields 
are not at all randomly tangled, but rather organized in folds 
within which the field remains straight up to the scale of the flow 
and reverses direction at the resistive 
scale \cite{Ott_review,Chertkov_etal_dynamo,SCMM_folding,SCTMM_stokes}.  
One immediate implication of the folded field structure is the criterion 
for the onset of nonlinearity. For incompressible MHD, back reaction 
is controlled by the Lorentz tension force~$\vB\cdot\nabla\vB\sim\kpar B^2$. 
This quantity 
depends on the parallel gradient of the field and does not know about 
direction reversals ($\kpar\sim\kd$ \cite{SCMM_folding}). 
Balancing~$\vB\cdot\nabla\vB\sim\vu\cdot\nabla\vu$, 
we find that the back reaction is important when the magnetic energy 
becomes comparable to the energy of the viscous-scale eddies. 
Clearly, some form of nonlinear suppression of stretching 
motions at the viscous scale must then occur. However, the eddies at 
larger scales are still more energetic than the magnetic field 
and continue to stretch it at their (slower) 
turnover rate. When the field energy reaches the energy of these 
eddies, they are also suppressed and it is the turn of yet larger 
and slower eddies to exert dominant stretching. 
The folded structure is preserved with folds elongating to 
the size~$\ls$ of the dominant stretching eddy. 
The key question is whether $\ls$ can increase all the way 
to the outer scale or stabilizes just above the viscous 
scale \cite{SCHMM_ssim}. 

The nonlinear suppression of stretching motions does not 
mean complete elimination of all turbulence: only the 
$\vb\vb:\nabla\vu$ component of the velocity-gradient tensor 
leads to work being done against the Lorentz force 
and, therefore, must be suppressed. It is then natural 
to expect a local anisotropization of the velocity field. 
In this Letter, we demonstrate how a simple model accounting 
for this nonlinearly induced local anisotropy can produce 
solutions that are in remarkably good agreement with 
numerically observed magnetic-energy spectra. 

The idea is to use the standard Kazantsev \cite{Kazantsev} 
model velocity, Gaussian and white in time, 
$\la u^i(t,\vx)u^j(t',\vx')\ra = \delta(t-t')\kappa^{ij}(\vx-\vx')$, 
but let $\kappa^{ij}$ depend on 
the local direction of the magnetic field, $\vb=\vB/B$. 
In the Lagrangian frame (with local rotation transformed out), 
$\vb$ orients itself along the stretching Lyapunov 
direction of the flow, which stabilizes 
exponentially in time \cite{Goldhirsch_Sulem_Orszag}. 
Therefore, in this frame, $\vb$ can be assumed to vary slowly with time.
In the presence of one preferred direction defined by~$\bb^i\bb^j$, 
the velocity correlator in $\vk$ space has the following 
form
\bea
\nonumber
\kappa^{ij}(\vk) &=& \kapI(k,|\mu|)\bl(\delta^{ij} - \nn_i\nn_j\br) 
+ \kapA(k,|\mu|)\bl(\bb^i\bb^j\br.\\
&&\bl. +\,\,\mu^2\nn_i\nn_j -\mu\bb^i\nn_j - \mu\nn_i\bb^j\br),
\label{kapij_def}
\eea
where $\vn=\vk/k$, $\mu=\vn\cdot\vb$.
Let us ignore the spatial dependence of all quantities that 
vary at the flow scale and slower. 
The velocity will only enter via its 
gradient $u^i_j\equiv\dd_j u^i$, which is now a function of time 
only with statistics 
$\la u^i_m(t)u^j_n(t')\ra = \delta(t-t')
\int\diff^3 k\,k_n k_m\kappa^{ij}(\vk)$. 
We can assume that $\vb$ also depends on time only, 
because it will always enter  via the tensor~$\bb^i\bb^j$, 
which varies at the scale of the flow  
(because of the folded structure of the magnetic field, 
the field's curvature is very small \cite{SCMM_folding,SCTMM_stokes}, 
so the fast spatial variation of $\vb$ is limited to sign reversals
and cancels in~$\bb^i\bb^j$).
With these assumptions, the solution to \eqref{ind_eq} can be 
written as (cf.~\cite{Zeldovich_etal_linear,Chertkov_etal_dynamo})
\bea
\label{Zeld_ansatz}
\vB(t,\vx) = \vb(t)\int\diff^3 k_0\,
\tB(t,\vk_0) e^{i\vx\cdot\tvk(t,\vk_0)},
\eea
where $\tvk(0,\vk_0)=\vk_0$ and 
\bea
\label{B_eq}
\dd_t\tB &=& \bb^i\bb^m u^i_m\tB - \eta \tk^2\tB,\\
\dd_t \tk_m &=& -u^i_m \tk_i,\\
\dd_t \bb^i &=& \bb^m u^i_m - \bb^l\bb^m u^l_m\bb^i.
\label{b_eq}
\eea
\Eqsdash{B_eq}{b_eq} are a modification of the 
so-called zero-dimensional model of the 
dynamo \cite{Gruzinov_Cowley_Sudan}. 
A closed equation can be obtained for the 
joint PDF of $\tB$, $\tvk$, and $\vb$,
${\cal P}(\tB,\tvk,\vb) = 
\delta(|\vb|^2-1)\delta(\vb\cdot\tvk)(4\pi^2 \tk)^{-1}P(\tB,\tk)$,
via an averaging procedure analogous to, e.g., the 
one in Ref.~\cite{SCMM_folding}. The magnetic-energy 
spectrum $M(k) = (1/2)\int_0^\infty\diff B\,B^2 P(B,k)$ 
is then found to satisfy 
\bea
\nonumber
\dd_t M &=& {1\over8}\,\gone{\dd\over\dd k}
\biggl[(1+2\spar)k^2{\dd M\over\dd k}\biggr.\\
\nonumber
&&\biggl.-\,\,(1+4\sperp+10\spar)k M\biggr]\\ 
&&+\,\,2(\sperp+\spar)\gone M - 2\eta k^2 M,
\label{FPEq}
\eea
where 
$\gone = \int\diff^3 k\,\kperp^2\kapperp$, 
$\sperp = (1/\gone)\int\diff^3 k\,\kpar^2\kapperp$,
$\spar = (1/\gone)\int\diff^3 k\,\kpar^2\kappar$, 
$\kperp = k(1-\mu^2)^{1/2}$, 
$\kpar = k\mu$, 
$\kapperp = (1/2)\bl(\delta^{ij} - \bb^i\bb^j\br)\kappa^{ij}$,
$\kappar = (1/2)\bb^i\bb^j\kappa^{ij}$, 
and $\kappa^{ij}$ is defined in \eqref{kapij_def}.
In the isotropic case, $\kapI=\kapI(k)$, $\kapA=0$, 
which gives $\sperp=2/3$, $\spar=1/6$. \Eqref{FPEq} then reduces 
to the standard equation for the magnetic-energy spectrum 
in the kinematic dynamo \cite{Kazantsev,KA}.
With a zero-flux boundary condition imposed at low~$k$ \cite{SCHMM_ssim}, 
\eqref{FPEq} has an eigenfunction (in the limit $\eta\to+0$)
\bea
\label{M_efn}
M(k) \simeq k^s e^{\gamma t} K_0(k/\kres), 
\eea
where 
$K_0$~is the Macdonald function, 
$\kres=\bl[(1+2\spar)\gone/16\eta\br]^{1/2}$,
$s=2(\sperp+2\spar)/(1+2\spar)$, and 
$\gamma=(\gone/8)\bl[16(\sperp+\spar)-(1+2\sperp+6\spar)^2/(1+2\spar)\br]$.
As magnetic back-reaction makes velocity more anisotropic, 
the values of~$\sperp$, $\spar$ drop compared to the isotropic case, 
and so does the growth rate~$\gamma$ ---
until the dynamo is shut down (for a purely 
two-dimensional velocity, $\sperp=\spar=0$ and 
$\gamma=-\gone/8$). 
Thus, {\em saturation can be achieved purely by anisotropizing 
the statistics of the velocity field.} 

How do we make connection from a theory based on the 
$\delta$-correlated model velocity to the real turbulence, 
which has a finite correlation time? The simplest prescription 
is to get finite expressions for equal-time velocity 
correlators by replacing the $\delta$~function by $1/\tcorr$:
$\la u^i(\vk)u^{j*}(\vk)\ra\equiv I^{ij}(\vk)=\tcorr^{-1}\kappa^{ij}(\vk)$. 
We take the correlation time $\tcorr$ of a given type of motions 
to be their ``turnover time'': 
defining $\Iperp$ and $\Ipar$ analogously to 
$\kapperp$ and $\kappar$, 
we write $\kperp^2\kapperp = \Cone\gone^{-1}\kperp^2\Iperp$, 
$\kpar^2\kapperp = \Cperp(\sperp\gone)^{-1}\Iperp$, 
and $\kpar^2\kappar = \Cpar(\spar\gone)^{-1}\Ipar$, 
where $\Cone$, $\Cperp$, $\Cpar$ are adjustable constants. 
Then 
$\gone = \bl[\Cone\int\diff^3 k\,\kperp^2\Iperp\br]^{1/2}$, 
$\sperp = 
\bl[(2/3)\int\diff^3 k\,\kpar^2\Iperp\bl/\int\diff^3 k\,\kperp^2\Iperp\br.\br]^{1/2}$,
and
$\spar  = 
\bl[(1/6)\int\diff^3 k\,\kpar^2\Ipar\bl/\int\diff^3 k\,\kperp^2\Iperp\br.\br]^{1/2}$, where  
we have set $\Cperp=(2/3)\Cone$, $\Cpar=(1/6)\Cone$ to ensure 
that $\sperp=2/3$ and $\spar=1/6$ in the isotropic case. 

In order to model gradual anisotropization of the velocity 
statistics by the back reaction, 
we define the stretching wave number~$\ks(t)$ such that 
the total magnetic energy~$W(t)$ at time~$t$ is 
equal to the energy of the 
hydrodynamic eddies at~$k>\ks$ (before they feel 
the nonlinearity). We assume that the eddies at $k<\ks$ 
remain isotropic (unaffected by back reaction), while 
those at $k>\ks$ are two-dimensionalized. 
Specifically, for $\kf<k<\ks(t)$, let 
\bea
4\pi k^2\Ii(k,|\mu|) = E(k),\quad 
\Ia(k,|\mu|)=0,
\eea 
while for $\ks(t)<k<\kd$, 
\bea
\label{Ii_2D}
4\pi k^2\Ii(k,|\mu|) &=& 2\rK E(k)\,\delta(\mu),\\
\label{Ia_2D}
4\pi k^2\Ia(k,|\mu|) &=& 2\Ea(k)\,\delta(\mu).
\eea
Here $\ks(t)$ is defined by 
$c_2\int_{\ks(t)}^{\kd} E(k) = W(t)$, 
$\Ii$ and $\Ia$ are coefficients of 
$I^{ij}$ analogous to $\kapI$ and $\kapA$ [\eqref{kapij_def}], 
$\kf$ and $\kd$ are the forcing and viscous wave numbers, 
$c_2$ and $\rK$ are adjustable parameters. 
We take $E(k)=\CK\epsilon^{2/3}k^{-5/3}$ (with $\CK=1.5$) 
for $k\in[\kf,\kd]$. The specific form of~$E(k)$ will only affect 
details of the transient time evolution, not the saturated state. 
$\Ea(k)$ will not figure in what follows, because 
it multiplies~$\mu\delta(\mu)$ in all relevant expressions. 
Coefficients in \eqref{FPEq} now depend on~$W(t)$: 
a straightforward calculation gives 
\bea
\gone(t) &=& 
{6\over5}\,\gKA\biggl[1-{1\over(1+\Wo/\Wd)^2}\biggr]^{-1/2} 
\bl[\Gamma(t)\br]^{1/2},
\label{gone_model}\\
\nonumber
\sperp(t) &=& 4\spar(t) =  
{2\over3}\,\biggl[ {1\over(1+W(t)/\Wd)^2}\biggr.\\
 &&\biggl.-\,\, {1\over(1+\Wo/\Wd)^2} \biggr]^{1/2} 
\bl[\Gamma(t)\br]^{-1/2},
\label{s_model}\\
\nonumber
\Gamma(t) &=&
{1\over(1+W(t)/\Wd)^2} - {1\over(1+\Wo/\Wd)^2}\\ 
\nonumber
&&+\,\, {5\over4}\,\rK\,\biggl[ 1 - {1\over(1+W(t)/\Wd)^2}\biggr],
\eea
where 
$\gKA=c_1\bl[\int_{\kf}^{\kd}\diff k\,k^2 E(k)\br]^{1/2}$, 
$c_1=\bl[(5/18)\Cone\br]^{1/2}$, 
the viscous-eddy energy is 
$\Wd/c_2=(3/2)\CK\epsilon^{2/3}\kd^{-2/3}$, 
and the total energy of the velocity field (before suppression) 
is $\Wo/c_2=\int_{\kf}^{\kd}\diff k\,E(k)$. 
\Eqsdash{gone_model}{s_model} 
represent a generalization of the model first introduced in 
Ref.~\cite{SCHMM_ssim} and reduce to it when $\rK=0$. 
They include the effect of quasi-2D mixing 
of the folded magnetic fields by eddies whose stretching 
component has been suppressed. The spectrum of these mixing 
motions is modelled by \eqref{Ii_2D}, where $\rK$ parametrizes 
the strength of the mixing relative to 
the original unsuppressed 3D turbulence. 

\begin{figure}[t]
\centerline{\psfig{file=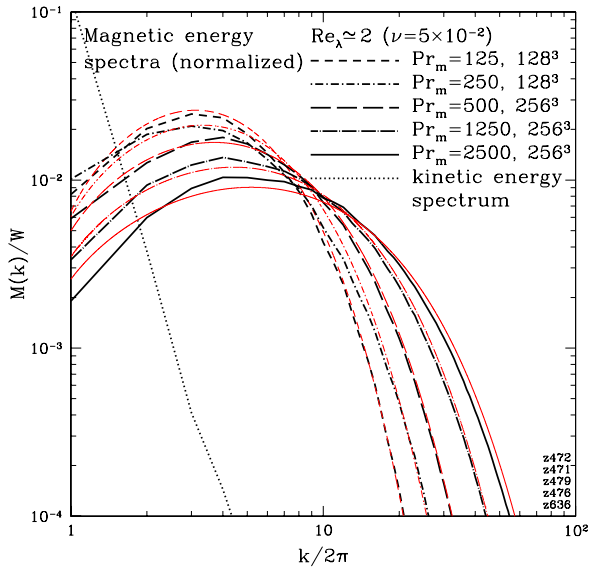,width=8cm}}
\caption{\label{fig_stokes} 
The bold lines are the normalized saturated 
energy spectra for simulations in the viscosity-dominated 
regime \cite{SCTMM_stokes}. 
The thin lines are the spectra predicted by our model.}
\end{figure}

The behavior of our model is easy to predict. 
The kinematic growth stage [$\gone=(6/5)\gKA$, 
$\sperp=2/3$, $\spar=1/6$, and $s=3/2$, $\gamma=(3/4)\gKA$ 
in \eqref{M_efn}] lasts until the total magnetic energy 
reaches the energy of the viscous-scale eddies, 
$W\sim\Wd$. After that, the velocity is gradually anisotropized, 
stretching is weakened, but mixing continues at $k>\ks(t)$. 
A steady solution is reached as soon as $\sperp$ and $\spar$ 
have decreased enough to render $\gamma=0$ in \eqref{M_efn}. 
This gives $\sperp=4\spar\simeq0.078$. The corresponding 
spectral exponent in the interval $\kd\ll k\ll\kres$ is 
$s\simeq0.23$. 
This solution is valid in the limit 
$\kres\gg\kd$ ($\Pr\gg1$), but convergence in~$\Pr$ is 
only logarithmic. In practice, 
numerical solution of \eqref{FPEq} shows that a scale separation 
of many decades is required for the scaling to be 
discernible. This is not achievable in 
direct numerical simulations. We have, therefore, solved 
\eqref{FPEq} with the same parameters as 
those used in our simulations \cite{SCTMM_stokes}. 
There are three adjustable constants: $c_1$, $c_2$, and $\rK$. 
The solution does not, however, depend very strongly on them: 
$c_2$ is irrelevant as it amounts to overall rescaling of energy,
$\rK$ has to vary by an order of magnitude to cause significant change, 
and even $c_1$ (which affects the value of $\kres$) does not require 
very fine tuning. 
We have compared the model solutions for the {\em same} fixed values 
$c_1=1/3$ and $\rK=4/5$ with the (normalized) spectra 
obtained in numerical simulations. A sequence of runs with 
large $\Pr$ and low $\Re$ (the so-called viscosity-dominated 
limit: a necessary compromise at current 
resolutions \cite{SCTMM_stokes}) is well 
fitted by our model in both kinematic (not shown) 
and nonlinear (\figref{fig_stokes}) regimes 
(except at $k/2\pi=1,~2$, where finite-box effects are important). 
Note that the velocity field in these runs is random 
[because of random forcing, \eqref{NSEq}], but, unlike 
in real turbulence, spatially smooth and one-scale. 

It is extraordinary that our minimal model 
has reproduced non-asymptotic 
numerical spectra so well. We do not claim 
that it constitutes a quantitative theory of nonlinear dynamo. 
It does, however, provide a simple demonstration that 
the available numerical data is consistent with magnetic-energy 
spectra exhibiting a very flat positive spectral exponent 
in the interval $\kd\ll k\ll\kres$ if sufficiently 
large scale separations were resolved.

It is clear that the viscosity-dominated simulations (low~$\Re$) 
are described very well by our model. 
The case $\Re\gg1$, $\Pr\gg1$ is much harder to tackle. 
If mixing by velocities at $k\in[\ks,\kd]$ remains 
efficient [as implied by our 2D 
approximation~\exsdash{Ii_2D}{Ia_2D}], 
then $\ks$ stabilizes at a value~$\sim\kd$ 
and saturated magnetic energy scales with $\Re$ 
as the energy of the viscous eddies, $\Bsq\sim\Re^{-1/2}\usq$. 
This outcome does not appear to be borne out by 
the available numerical evidence, which rather suggests 
$\Bsq\lesssim\usq$ \cite{Haugen_Brandenburg_Dobler} 
(though limited resolutions preclude a definitive statement). 
In our runs with 
$\Pr=10$ and Taylor-microscale Reynolds number 
$\Rel\simeq45$ ($\Re\sim100$)
and with $\Pr=1$, $\Rel\simeq155$ ($\Re\sim400$), 
our model in its present form 
overestimates the magnetic energy at large~$k$,
but underestimates it at low~$k$
(\figref{fig_AB}): an indication of too much mixing in 
the model \footnote{It is fair to acknowledge that the validity 
of our model for these runs is also questionable because 
$\Pr$ is not large.}.
Indeed, when $\Re\gg1$, the nature of the anisotropized 
velocities in the interval $[\ks,\kd]$ 
can be very different from the interchange-like motions 
that give the 2D mixing in the viscosity-dominated case. 
In Ref.~\cite{SCHMM_ssim}, we argued that the interval 
$[\ks,\kd]$ is populated 
by Alfv\'en waves that propagate along the folds. 
The saturated spectrum is then the result of a 
superposition of waves and folds 
(which accounts for the large amount of small-scale magnetic energy). 
Since the Alfv\'en waves are dissipated by viscosity, 
they can only exist if the stretching scale becomes much 
larger than the viscous scale: possibly as large 
as the outer scale ($\ks\sim\kf$, cf.~\cite{SCTMM_stokes}). 
This is only allowed if 
the waves do not mix magnetic field as efficiently 
as the interchange motions do. 
For our model, the required modification 
would be that the mixing rate $\gone$ should decrease with $\ks$. 
The dynamo saturation would then be due to 
a balance between stretching and mixing by partially 
anisotropized motions {\em at the stretching scale}.

Detecting Alfv\'en waves along folds 
is a challenge for future numerical work. 
The main conclusion of the present study is that the nonlinear dynamo 
in a random one-scale flow can be described by a simple 
model where saturation is achieved via partial anisotropization 
of the ambient velocity, a result quantitatively supported 
by agreement with direct numerical simulations. 

\begin{acknowledgments}
Our work was supported by grants from 
PPARC (PPA/G/S/2002/00075), 
EPSRC (GR/R55344/01), 
UKAEA (QS06992),
NSF (AST~00-98670). 
Simulations were done at UKAFF (Leicester) 
and NCSA (Illinois). 
\end{acknowledgments}

\bibliography{scthmm_PRL}

\end{document}